\documentclass{PoS}
\usepackage{amsmath,amssymb,graphicx}

\title{M\"obius Algorithm for Domain Wall and GapDW Fermions }

\ShortTitle{M\"obius Algorithm for Domain Wall and GapDW Fermions }

\author{\speaker{Richard C. Brower}%
\\
        Boston University\\
        E-mail: \email{brower@bu.edu}}

\author{Ronald Babich\\
        Boston University\\
        E-mail: \email{rbabich@bu.edu}}
\author{Kostas Orginos\\
        William \& Mary University\\
        E-mail: \email{kostas@wm.edu}}
\author{Claudio Rebbi\\
        Boston University\\
        E-mail: \email{rebbi@bu.edu}}
\author{David Schaich\\
        Boston University\\
        E-mail: \email{schaich@bu.edu}}
\author{Pavlos Vranas\\
        Lawrence Livermore National Laboratory\\
        E-mail: \email{vranas2@llnl.gov}}

      \abstract{The M\"obius domain wall action~\cite{Brower:2004xi}
        is a generalization of Shamir's action, which gives exactly
        the same overlap fermion lattice action as the separation
        ($L_s$ ) between the domain walls is taken to infinity. The
        performance advantages of the algorithm are presented for
        small ensembles of quenched, full QCD domain wall and Gap
        domain wall lattices~\cite{Vranas:2006zk}. In particular, it
        is shown that at the larger lattice spacings relevant to
        current dynamical simulations M\"obius fermions work well
        together with GapDWF, reducing $L_s$ by more than a factor of
        two. It is noted that there is a precise map between the
        domain wall and effective overlap action at finite quark mass
        including finite $L_s$ chiral violations so that the
        Ward-Takahashi identities for the axial and vector currents
        are exactly equivalent in the two formulations. }

\FullConference{The XXVI International Symposium on Lattice Field Theory \\
                 July 14 - 19, 2008\\
                 Williamsburg, Virginia, USA}

\begin{document}

\section{Introduction}

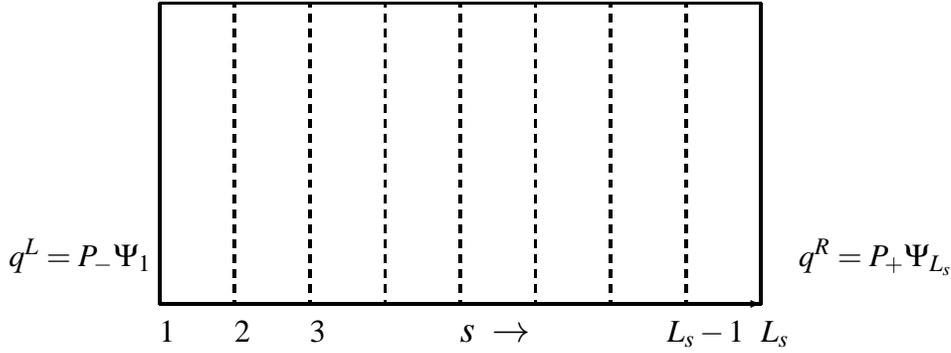
\begin{figure}[ht!]
\begin{center}
\setlength{\unitlength}{0.5 mm}
\begin{picture}(220,100)
\linethickness{.35mm}
\put(20,10){\vector(1,0){160}}
\multiput(40,10)(40,0){4}{\dashbox{2}(20,80)}
\multiput(20,10)(20,0){1}{\line(0,1){80}}
\multiput(180,10)(20,0){1}{\line(0,1){80}}
\put(20,90){\line(1,0){160}}
\put(190,20){\large $q^R = P_+ \Psi_{L_s}$}
\put(-20,20){\large $q^L = P_- \Psi_1$}  
\put(20,0){\large $ 1$}  
\put(40,0){\large $ 2$}  
\put(60,0){\large $ 3$}  
\put(100,0){\Large $s \; \rightarrow $}  
\put(155,0){\large $L_s - 1$}  
\put(180,0){\large$ L_s$}  
\end{picture}
\caption{Domain wall convention with left/right chiral mode at $s = 1$ and $s =  L_s$ respectively approximated by 3-branes separated by a distance $a_5   L_s$. The Pauli Villars ($m=1$) and the zero mass Dirac ($m=0$) operators obey anti-periodic and Dirichlet boundary conditions respectively.}
\label{Fig:domainwall}
\end{center}
\end{figure}

Domain wall fermions provide an efficient and rigorous implementation
of chiral symmetry in lattice field theory at finite lattice
spacing. Following the original ideas of Kaplan and Shamir one
introduces two 4-d domain walls (or 3 branes) separated by $L_s$
lattice sites in a 5th dimension. The 5-d domain wall action,
\begin{equation} S_{DW} = \sum_{x,s}[\Psi_{x,s} (D_{DW}(m) \Psi)_{x,s} + \Phi_{x,s}
(D_{DW}(1) \Phi)_{x,s}]  \; ,
\end{equation} 
contains 5-d Wilson fermion ($\Psi_{x,s}$) and Pauli-Villars
pseudo-fermion  ($\Phi_{x,s}$) fields which enjoy a ``kinematical''
super symmetry broken only by the boundary conditions on the 3-branes
(see Fig.~\ref{Fig:domainwall}).
The result is an effective 4-d action with low mass states below the
cut-off representing a left(right) chiral fermion on each wall at $s =
1(L_s)$ respectively in the limit of infinite separation between the
walls. After a long and interesting history of competing methods, it
is now realized that the effective 4-d theory at $L_s = \infty$ is
equivalent to one based on Neuberger's overlap operator, with
\begin{equation}
S_{ov}=  \sum_{xy} \psi_x D_{ov,xy}(m)\psi_{y}  \equiv
 \psi \Big[ \frac{1+m}{2} + \frac{1-m}{2} \gamma_5 
\epsilon[H]\Big]  \psi 
\end{equation}
and an appropriate Dirac ``Hamiltonian'' in the sign function
$\epsilon[H]$. The two actions lead to equivalent matrix elements
$D^{-1}_{ov,xy} (m) \equiv \langle \psi_x
\overline \psi_y \rangle_{ov} = \langle q_x \overline q_y \rangle_{DW}
$, where the fields $q, \bar q$, shown in Fig.~\ref{Fig:domainwall}, are
mathematically defined in Sec.~\ref{sec:discussion}.

For zero mass quarks, $D_{ov}(0)$ obeys 
the Ginsparg-Wilson relation,
\begin{equation}
\gamma_5 D_{ov}(0) + D_{ov}(0) \gamma_5 = 2D_{ov}(0) \gamma_5 D_{ov}(0) \; ,
\end{equation}
or equivalently the anti-commutator, $\{ \gamma_5, D^{-1}_{ov}(0) \} =
2 \gamma_5$, which guarantees exact chiral symmetry at finite lattice spacing
and zero quark mass, $m_f = m/(1-m)$.  Since all implementations give
solutions to the GW relation, the debate on the virtues of overlap vs. domain
wall fermions is essentially algorithmic in nature. For domain wall
algorithms, practical considerations demand that the $L_s \rightarrow
\infty$ limit be approximated by modest values: $L_s = O(10)$.  This
is not a trivial requirement. Finite $L_s$ causes a residual breaking
to the GW relation as measured by the difference operator,
\begin{equation}
2 \gamma_5 \Delta_{L_s}[H] =  \gamma_5 D_{ov}(0) + D_{ov}(0) \gamma_5 - 2 D_{ov}(0) \gamma_5 D_{ov}(0) \; .
\label{eq:Delta}
\end{equation}

The conventional criterion for estimating this violation of chiral
symmetry is to compare the magnitude of ``residual mass'',
\begin{equation}
m_{res} = \frac{Tr[D^{\dagger -1}_{ov} \; \Delta_{L_s}[H] \; D^{ -1}_{ov}]}{Tr[D^{\dagger -1}_{ov} D^{ -1}_{ov}]} \, 
\end{equation}
relative to the explicit quark mass $m_f = m/(1-m)$. 
As emphasized by Sharpe~\cite{Sharpe:2007yd}, at current lattices
spacings a residual mass $O(10^{-3})$ is adequate but some quantities
require an order of magnitude smaller residual mass. Moreover recent
applications to finite temperature and ${\cal N} =1$ SUSY QCD have
required much larger values of $L_s = O(100)$. Better methods are
needed to reduce $m_{res}$ at reasonable values of $L_s$.

Within this framework there still remains a large space of options
for the lattice partition function,
\begin{equation}
Z[U]  = \int  {\cal D} \overline \psi {\cal D} \psi \; e^{\; \textstyle \beta Tr[ U_P] + S_{improved}[U] +  \overline \psi D_{ov}(m) [U]\psi }
\end{equation}
by improving the gauge action and/or the approximation to the overlap
operator. Here we re-examine the M\"obius formulation of the domain
wall algorithm~\cite{Brower:2004xi}, verifying that very substantial
improvements can be made in the convergence rate to the exact chiral
fermion at $L_s = \infty$. In addition there is a strong feedback
between improved gauge and fermionic algorithms. In particular we
point out when the so called ``Gap domain wall'' modification of the
gauge action combines nicely with the M\"obius fermion action to give
multiplicative improvements --- each reducing the residual mass by
separate orders of magnitude. 


\section{M\"obius Recipe}

The simplest kernel for the overlap algorithm is the Wilson Hamiltonian
operator, $H = \gamma_5 D^{Wilson}(M_5)$, where
\begin{equation}
D^{Wilson}_{xy}(M_5) =  (4+M_5) \delta_{x,y} - 
 \frac{1}{2} \Bigl[  (1 - \gamma_\mu) U_{x,x+\mu}\delta_{x+\mu,y} 
+  (1 + \gamma_\mu) U_{x,x+\mu}^\dagger \delta_{x,y+\mu} \Bigr] \; , 
\label{eq:D^{Wilson}} 
\end{equation}
with a negative mass parameter $M_5 \in [-1,-2]$. However for the 
domain wall algorithm, the simplest implementation is
the Shamir form, $H = \gamma_5 D^{Shamir}(M_5) $,
\begin{equation}
D^{Shamir}(M_5) = \frac{a_5 D^{Wilson}(M_5)}{2 + a_5 D^{Wilson}(M_5)}  \; . 
\label{eq:Shamir}
\end{equation}
The M\"obius form is a  real 3 parameter  M\"obius transform of the Wilson kernel interpolating between both of these,
\begin{equation}
D^{Moebius}(M_5) =  \frac{(b_5 + c_5)D^{Wilson}(M_5)}{2 +  (b_5 - c_5) D^{Wilson}(M_5)}  \equiv \alpha D^{Shamir}(M_5) \; .
\end{equation}
Relative to Shamir's kernel this introduces 1 new ``scaling'' parameter $\alpha = (b_5 + c_5)/a_5$ at fixed $a_5 = b_5 - c_5$. Due to the scale invariance of the sign function, $\epsilon[\alpha \lambda] = \epsilon[\lambda]$, this does not change the $L_s = \infty$ chiral lattice action.  {\bf Consequently at fixed $a_5, M_5$ the   M\"obius rescaling should be regarded as an improved algorithm for the same chiral action,
optimized by   choosing $\alpha(L_s)$ to minimize $m_{res}$ at finite $L_s$.}

Why is this freedom to rescale $H$ desirable? The difficulty
with the domain wall approach is that the resultant polar
approximation to the sign function, 
\begin{equation}
  \epsilon_{L_s} [H] = \frac{(1+H)^{L_s}  - (1-H)^{L_s}}
                             {(1+H)^{L_s}  + (1-H)^{L_s}}   =   \tanh[- (L_s/2)\log T] \; ,
\label{eq:ApproxEPS}
\end{equation}
is exponentially convergent only for eigenvalues $\lambda$ of $H$
inside the interval: $\log(|\lambda|) \in [1/L_s,L_s]$. So the
advantage of rescaling at finite $L_s$ is to use this interval more
efficiently by shifting the spectrum $\log(\lambda) \rightarrow
\log(\alpha) + \log(\lambda)$.  Other approaches to improving the
polar approximation that have been suggested include explicit
projection of a finite set of eigenvalues at small $\lambda$ and/or
suppressing the number of small eigenvalues by changing the gauge
action. Indeed these may be combined together with M\"obius fermions
to gain additional advantage as illustrated here by the M\"obius
rescaling of the GapDW lattices.  Next we explain how the domain wall
implementation of this rescaling naturally involves the two parameters,
$b_5, c_5$.

The M\"obius generalization of Shamir merely requires that the Wilson
kernel be included in the 5th dimensional hopping term,
\begin{eqnarray}
D^{DW}(m)_{s,s'}  &=& 
 D^{(s)}_- \, P_+ \,\delta_{s,s'+1}  +  D^{(s)}_+ \, \delta_{s,s'} 
+   D^{(s)}_- \, P_-  \, \delta_{s,s'-1}  \\
 &-&  m \, D^{(1)}_- P_+ \, \delta_{s,1}\delta_{s',L_s}  - m \,  D^{(L_s)}_- P_-
\, \delta_{s,L_s}\delta_{s',1} \nonumber 
\label{eq:DWaction}
\end{eqnarray}
with $P_\pm = \frac{1}{2}(1 \pm \gamma_5)$ and $D^{(s)}_+ = b_5(s)
D^{Wilson}(M_5) +1$, $D^{(s)}_- = c_5(s) D^{Wilson}(M_5) -1$ with
$s,s' = 1, 2, \cdots L_s$ or in $L_s \times L_s$ matrix notation.  For
the rescaling example discussed we take $b_5(s) + c_5(s) =
 \alpha a_5 \; ,\; b_5(s) - c_5(s) = a_5$, so the s-dependence for $D^{(s)}_\pm$
can be dropped, however we have included it so that the M\"obius class
includes other approaches such as the Zolotarev approximation or the
variable fields suggested by B\"ar, Narayanan, Neuberger and
Witzel~\cite{Bar:2007ew}.  In matrix notation:
\begin{equation}
D^{DW}(m) =
\begin{bmatrix}
D^{(1)}_+ & \quad D^{(1)}_- P_- & 0 &\cdots &  -mD^{(1)}_- P_+  \\
\quad D^{(2)}_- P_+ & D^{(2)}_+ & \quad D^{(2)}_- P_- &\cdots & 0  \\
0 & \quad D^{(3)}_- P_+ & D^{(3)}_+ & \cdots&   0  \\
\cdots & \cdots & \cdots & \cdots &  \cdots  \\
-mD^{(L_s)}_- P_- & 0 & 0  &  \cdots & D^{(L_s)}_+  \\
\end{bmatrix}
\end{equation}

\section{Performance Measures}

Fortunately the new off diagonal Wilson operators in the M\"obius domain wall action can be implemented with essentially no additional algorithmic complexity. The first step, suggested in Ref~\cite{Brower:2004xi}, is to replace 5-d red/black preconditioning by a 4-d checker board with no alternation of color along the 5th axis. The new form of the Schur complement solves analytically all interaction in the fifth dimension. The performance of 4-d versus 5-d red/black preconditioning, if anything, favors this construction.  Then with a simple gather operation for the three spinors in the $\hat \mu$ direction, the number of Wilson Dirac application per CG iteration is identical. A full comparison of performance on a range of lattices is impossible in this short talk. So we consider three examples with more to be presented in a future publication.

\begin{figure}[h!]
\begin{center}
\includegraphics[width=0.65\textwidth]{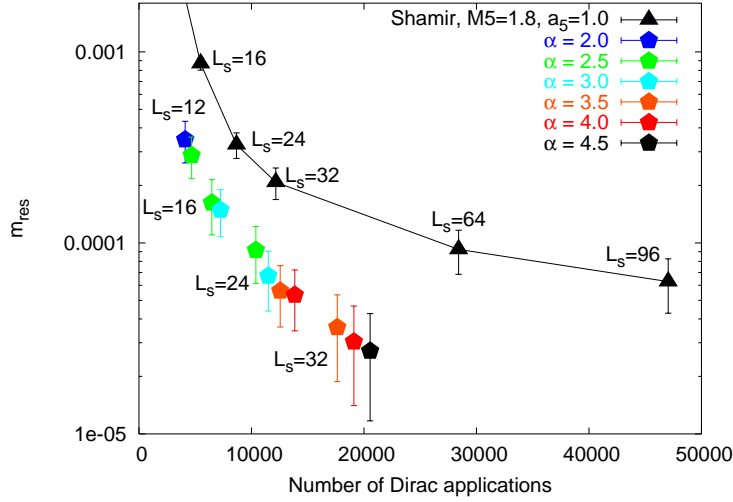}
\caption{ The M\"obius algorithm on pure gauge lattices compared with Shamir
($\alpha = 1$).}
\label{Fig:fig5}
\end{center}
\vskip -0.5cm
\end{figure}
{\bf Quenched  Lattices:} As demonstrated in the original proposal~\cite{Brower:2004xi}, the M\"obius formulation has the potential of an order of magnitude reduction of the explicit chiral symmetry breaking at fixed computational cost.  A comparison of the residual masses is given in Fig.~\ref{Fig:fig5} on $\beta = 6.0$ quenched lattices with $a^{-1} \simeq 2.1GeV$.  For these lattices the optimal rescaling satisfies the empirical form, $\alpha(L_s) \simeq 1+ L_s/8$. Note that for small residual masses the advantage of scaling is huge. 

\begin{figure}[h!]
\center{\includegraphics[width=0.65\textwidth]{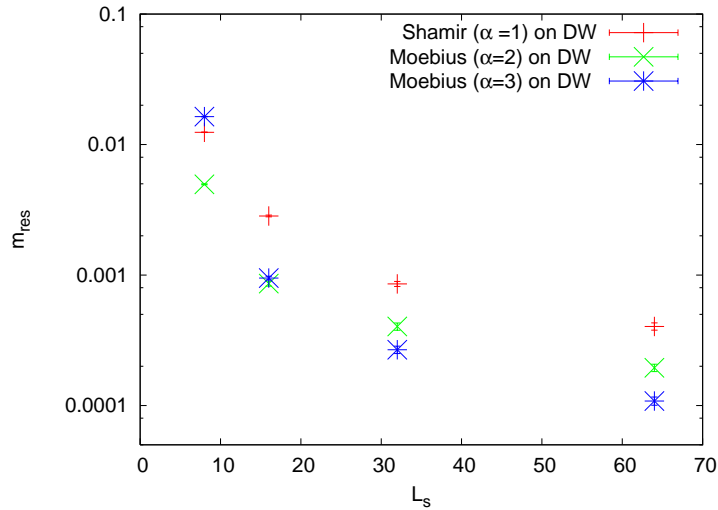}}
\label{Fig:dw}
\caption{ The M\"obius algorithm  on full QCD domain wall lattices compared to  Shamir ($\alpha =1$).}
\end{figure}

{\bf Domain Wall Lattices:} Test were also carried out on full domain wall lattices which generally exhibit worse convergence to small $m_{res}$. However as an example in Fig.~\ref{Fig:dw} the same comparison is made on a set of DWF $N_f = 2+1$ Iwasaki lattices~\cite{Allton:2008pn} at $\beta = 2.13$ with $m_s=0.04, m_l/m_s = 1/4$, $a^{-1} \simeq 1.7 GeV$.  Even without carefully tuning the rescaling parameter the advantage appears to be nearly as dramatic as for the quenched lattices. More thorough studies of this are underway.

\begin{figure}[h!]
\begin{center}
\includegraphics[width=0.65\textwidth]{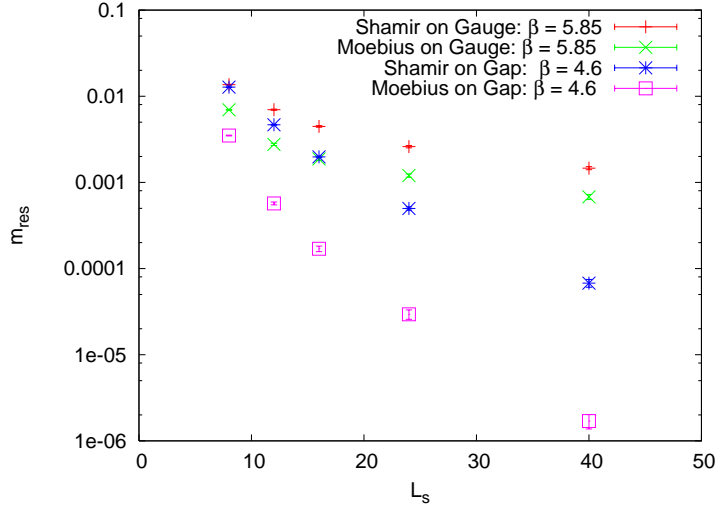}
\caption{ The M\"obius algorithm ($\alpha = 2$) on pure gauge and
Gap lattices vs Shamir ($\alpha = 1$).}
\label{Fig:test}
\end{center}
\vskip - 0.5cm
\end{figure}
{\bf Gapped Lattices:} A more radical suggestion to suppress small 
eigenvalues in $H$ was suggested by Vranas~\cite{Vranas:2006zk} which by adding a pair of Wilson fermions with mass $M_5 < 0$ induces an improvement term,
\begin{equation}
S_{Gap}[U_\mu(x)] = Tr[\log(D^{\dagger Wilson}(M_5) D^{Wilson}(M_5))] \; ,
\end{equation}
for the effective gauge action.  To test the locality of this action
one should recognize that the bilocal dependence on the gauge field is
measured by ``axial correlator''
\begin{equation}
\delta_{U_\mu(x)} \delta_{U_\nu(y)} S_{Gap}[U] =
\langle j^{5\mu}_{Gap}(x) j^{5\nu}_{Gap}(y) \rangle \sim \exp[ -|x-y|/\xi_\pi]
\end{equation}
whose long distance tail is the would be ``pion'' propagator for the
Gap fermions. Ref.~\cite{Vranas:2006zk} demonstrated that this
correlation is O(1) in lattice units as you approach the continuum
limit, roughly the same degree of locality in the overlap action itself. Fig.~\ref{Fig:test} compares  ($\alpha = 2$) vs Shamir ($\alpha = 1$)
at $M_5 = -1.8$, $m=0.02$ for pure gauge and Gapped lattices
both with  $a^{-1} = 1.4 Gev$.

\section{Discussion}
\label{sec:discussion}

Figs.~\ref{Fig:fig5}-~\ref{Fig:test} point to a general rule that a  M\"obius scaling transformation of Shamir by a factor of $\alpha > 1$ allows one to reduce $L_s$ and therefore the cost by a factor of $1/\alpha$ at fixed $m_{res}$.  This is a natural consequence of the scaling relation, $\Delta_{L_s}[\alpha \lambda] = \Delta_{\alpha L_s}[\lambda]$, for small eigenvalues, $|\lambda| \le O(1/L_s)$.  Tuning $\alpha >2$ for large $L_s$ gives additional savings. There appears to be an additional computational cost for $\alpha \ge 2$ on the order of $10\%$ due to increased condition number.  Alternatively we note at fixed $L_s = O(10)$, the M\"obius algorithm can reduce $m_{res}$ by an order of magnitude or more at fixed cost. The combined improvement in the M\"obius algorithm on Gapped lattices is multiplicative so here one might contemplate exploratory studies in beyond the standard model strong dynamics with $L_s$ in the range of 4 to 8. For high precision studies $m_{res}$ can be reduced below $10^{-5}$ for $L_s = 32$.  Both of these are attractive options.

In a subsequent publication more details on the efficiency and formal properties of the M\"obius algorithm will be provided.  We simply note here that a straight forward general formalism exists that allows all correlators as well as the Ward-Takahashi identities to be expressed independent of the detailed form of the M\"obius domain wall action.  For example
the application of LDU decomposition leads to the basic identity,
\begin{equation}
  [{\cal P}^{\dagger} \frac{1}{D^{DW}(m)}D^{DW}(1) {\cal P}]_{s's} = 
\begin{bmatrix}
D^{-1}_{ov}(m) & 0 & 0 & \cdots& 0\\
(1-m)\Delta^R_{2} D^{-1}_{ov}(m)  & 1 &0 &\cdots& 0\\ 
(1-m)\Delta^R_{3} D^{-1}_{ov}(m)  & 0 &1 &\cdots&0\\
\cdots & \cdots & \cdots & \cdots &  \cdots  \\
(1-m) \Delta^R_{L_s}  D^{-1}_{ov}(m)  & 0 & 0 &\cdots & 1
\end{bmatrix}_{s's}  \; ,
\label{eq:inverseDWFmatrix}
\end{equation}
where ${\cal P}_{s',s} = P_- \delta_{s',s} + P_+ \delta_{s',s+1}$
rotates the Right fermion to the Left wall in
Fig.~\ref{Fig:domainwall}.  It follows from this that
\begin{equation}
D^{-1}_{ov,xy} (m) \equiv \langle \psi_x
\overline \psi_y \rangle_{ov} = \langle q_x \overline q_y \rangle_{DW} \; ,
\end{equation}
defining the appropriate chiral domain wall fields on the boundary:
$ q_x = [{\cal P^\dagger} \Psi]_{x,1} \quad \mbox{and} \quad \overline q_x = [\overline \Psi D_{DW}(1) {\cal P} ]_{x,1} $ .
All the matrix elements are directly related to the transfer matrix, $T = (1 - H)/(1+H)$, through the partial left and right products~\footnote{For simplicity we have dropped the s-independence   of transfer matrix. The general case requires s-ordered product.},  $ \Delta^L_s = T^{-s}/( 1 + { T^{-L_s}})$ and $\Delta^R_{s+1} = T^{s-   L_s}/( 1 + T^{-L_s})$. Together
they give  the GW chiral breaking operator, $\Delta_{L_s}[H]= \Delta^L_{s} \Delta^R_{s+1}$, defined above in Eq.~\ref{eq:Delta}. Similar arguments leads to a general map between {\bf all} overlap and domain wall correlators, with only implicit reference to the M\"obius operator $D_{DW}(m)$. From this map, the vector and axial Ward-Takahashi identities must be identical for both overlap and domain wall actions at finite $L_s$, lattice spacing and finite volume.

In summary little change in the formalism or software is required to use the M\"obius algorithm, while providing a substantial improvement in performance. A very efficient code for the BlueGene has been written by Andrew Polchinski under the SciDAC software project and is readily available at the software links for USQCD: {\em http://usqcd.org}.

This work was supported in part by US DOE grant DE-FG02-91ER40676, NSF  grant DGE-0221680, NSF CCF-0728915, and the Jeffress Memorial Trust grant J-813.

\end{document}